\documentclass[sigconf,nonacm,10pt]{acmart}
\usepackage{amsmath}  

\usepackage{algorithmic}
\usepackage{graphicx}
\usepackage{textcomp}
\usepackage{subcaption}
\usepackage{braket}
\usepackage{url}
\usepackage{booktabs} 
\usepackage{soul,xspace,tabularx}
\usepackage{algorithmic}
\usepackage{algorithm}
\usepackage{array}
\usepackage{booktabs}
\usepackage{colortbl}

\newcommand{\eat}[1]{}

\def\BibTeX{{\rm B\kern-.05em{\sc i\kern-.025em b}\kern-.08em
    T\kern-.1667em\lower.7ex\hbox{E}\kern-.125emX}}
\begin{document}

\title{Benchmarking Quantum Data Center Architectures: A Resource and Topology Perspective}

\title{Benchmarking Quantum Data Center Architectures: A Performance and Scalability Perspective}

\author{Shahrooz Pouryousef}
\affiliation{
  \institution{Quantum Lab, Cisco Research}
  \country{USA}
}

\author{Eneet Kaur}
\affiliation{
  \institution{Quantum Lab, Cisco Research}
  \country{USA}
}

\author{Hassan Shapourian}
\affiliation{
  \institution{Quantum Lab, Cisco Research}
  \country{USA}
}

\author{Don Towsley}
\affiliation{
  \institution{University of Massachusetts Amherst, USA}
  \country{}
}

\author{Ramana Kompella}
\affiliation{
  \institution{Quantum Lab, Cisco Research}
  \country{USA}
}

\author{Reza Nejabati}
\affiliation{
  \institution{Quantum Lab, Cisco Research}
  \country{USA}
}




\begin{abstract}
Scalable distributed quantum computing (DQC) has motivated the design of multiple quantum data-center (QDC) architectures that overcome the limitations of single quantum processors through modular interconnection. While these architectures adopt fundamentally different design philosophies, their relative performance under realistic quantum hardware constraints remains poorly understood.

In this paper, we present a systematic benchmarking study of four representative QDC architectures—QFly, BCube, Clos, and Fat-Tree—quantifying their impact on distributed quantum circuit execution latency, resource contention, and scalability. Focusing on quantum-specific effects absent from classical data-center evaluations, we analyze how optical-loss-induced Einstein–Podolsky–Rosen (EPR) pair generation delays, coherence-limited entanglement retry windows, and contention from teleportation-based non-local gates shape end-to-end execution performance. Across diverse circuit workloads, we evaluate how architectural properties such as path diversity and path length, and shared BSM (Bell State Measurement) resources interact with optical-switch insertion loss and reconfiguration delay. Our results show that distributed quantum performance is jointly shaped by topology, scheduling policies, and physical-layer parameters, and that these factors interact in nontrivial ways. Together, these insights provide quantitative guidance for the design of scalable and high-performance quantum data-center architectures for DQC.

\end{abstract}



\maketitle


\section{Introduction}

Distributed Quantum Computing (DQC) has emerged as a promising approach for scaling quantum systems beyond the limits of single quantum processors by interconnecting multiple smaller Quantum Processing Units (QPUs) through quantum networks \cite{sakuma2024optical,cirac1999distributed,zhang2025switchqnet,shapourian2025quantum}. In DQC, circuits are partitioned across QPUs, and non-local operations are executed using entanglement-assisted primitives such as gate or state teleportation. Recent experimental demonstrations across multiple qubit platforms have validated the feasibility of such non-local operations \cite{main2025distributed,sinclair2025fault}, while compiler and orchestration frameworks based on lattice surgery and distributed surface codes for error correction have further advanced fault-tolerant distributed quantum computing \cite{watkins2024high,horsman2012surface,keskin2025lattice,cuomo2023optimized,zhang2025switchqnet}.

\begin{figure}
    \centering
    \includegraphics[scale=0.34]{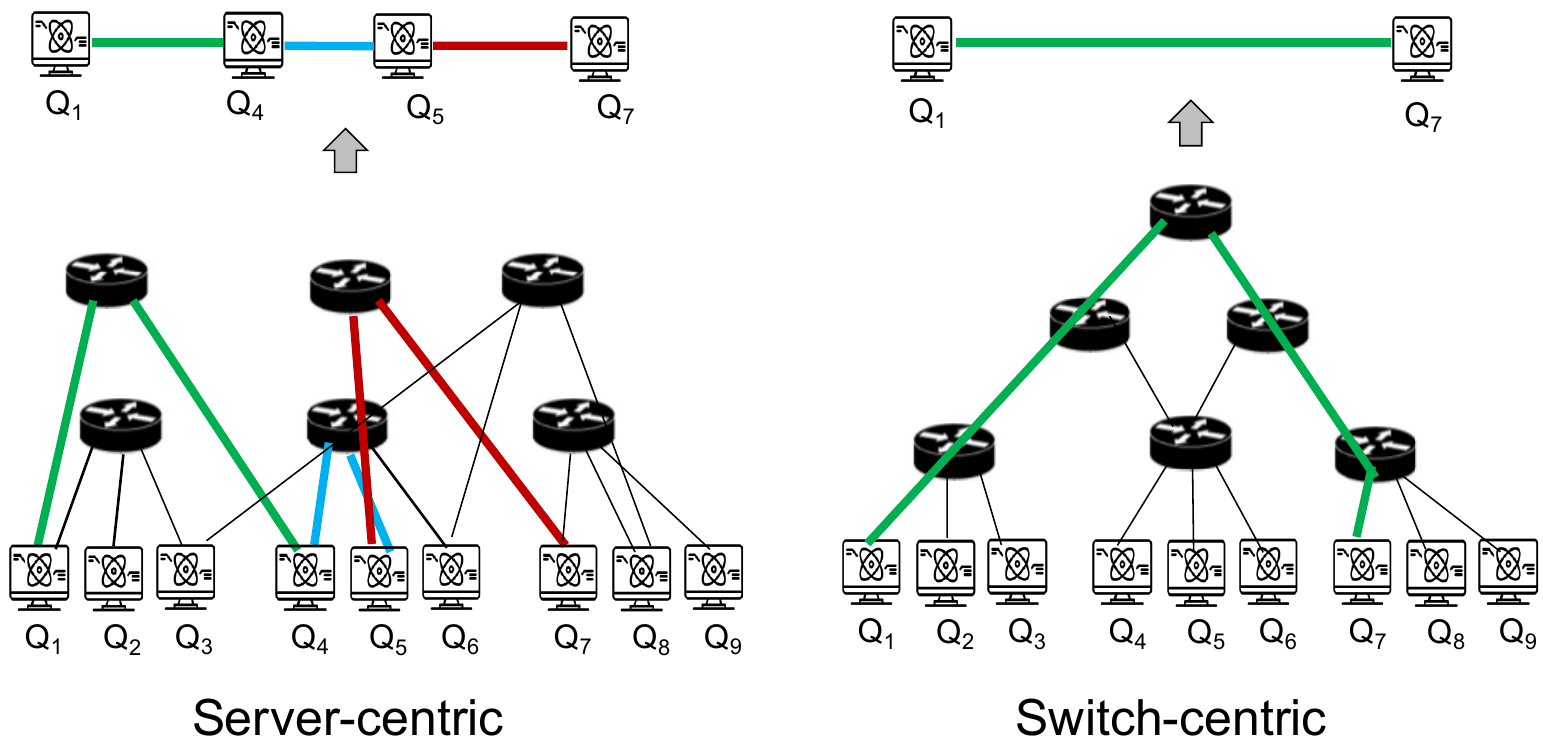}
    \caption{Entanglement routing in server-centric architectures using QPUs as repeaters versus switch-centric optical switching architectures.}
    \label{fig:QPUs_as_repeater_BCUBE}
\end{figure}

To support DQC at scale, several quantum data-center (QDC) architectures have been proposed, many inspired by classical data-center topologies but adapted to quantum networking constraints. These architectures emphasize different design goals, such as minimizing path length (e.g., Qfly \cite{sakuma2024optical}) or maximizing path diversity (e.g.,Fat-tree \cite{shapourian2025quantum}). In the classical domain, such topologies have been extensively benchmarked using metrics including throughput, fault tolerance, and capital efficiency \cite{popa2010cost,liu2013data,xia2016survey}. However, a comparable, cross-architectural evaluation remains largely absent in the quantum setting.

While both classical and quantum data-center benchmarking must consider factors beyond topology, QDC benchmarking introduces additional quantum-specific constraints that are absent in classical settings. In particular, distributed execution depends on stochastic entanglement generation subject to optical loss, coherence-limited retry windows for communication qubits, and teleportation-based non-local gates that contend for shared Bell-state measurement modules (BSMs) and detector resources. These effects interact nonlinearly with architectural properties, including switch hierarchy, path length between QPUs, available path diversity, and the distribution of shared photonic resources. As a result, improvements in individual hardware parameters—such as reduced switch insertion loss, faster reconfiguration of optical switches when establishing photonic paths between different QPU pairs, or increased communication-qubit availability—do not translate uniformly into system-level gains. Understanding when performance is dominated by topology, by resource contention, or by physical-layer limitations remains an open question.

Figure~\ref{fig:QPUs_as_repeater_BCUBE} illustrates how quantum-specific constraints arise from the physical realization of quantum interconnects and shape how entanglement is established between QPUs. In \emph{server-centric} architectures such as BCube, communication paths form repeater chains in which intermediate QPUs actively perform entanglement swapping, introducing and causing latency and noise to accumulate across multiple hops and making EPR generation sensitive to scheduling under finite coherence windows~\cite{pouryousef2024minimal,li2025optimising}. In contrast, \emph{switch-centric} architectures route entanglement through the optical switching fabric, where QPUs act only as endpoints and swapping is performed at BSM-capable switches along the path. In this figure, for both cases, we abstract each non-local operation as consuming a single \emph{virtual entanglement link} between the communicating QPUs, which aggregates physical losses and operational imperfections. 



In this work, we present a systematic benchmarking of four representative QDC architectures-QFly \cite{sakuma2024optical}, BCube \cite{shapourian2025quantum}, Clos \cite{shapourian2025quantum}, and Fat-Tree \cite{shapourian2025quantum}--under diverse circuit workloads and physical-layer conditions. Our evaluation explicitly examines how architectural design choices affect (i) end-to-end circuit execution latency under realistic entanglement-generation models, (ii) the number of switch hops incurred by non-local gates, and (iii) contention for BSM resources. We study multiple architectural variants to isolate the trade-offs between path length, contention, and scalability. Workloads span a range of circuit widths and depths, connectivity patterns, and two-qubit gate densities, enabling us to assess how algorithmic structure interacts with architectural and physical constraints.

Our study yields several architecture-level insights for designing scalable QDCs. First, performance is tightly coupled to how BSM resources are provisioned. Under a fixed
\emph{per-switch} BSM capacity model, architectures with a larger number of switches and greater path diversity—such as Clos and Fat-Tree—achieve lower latency by exposing a higher aggregate BSM pool and enabling more parallel
entanglement generation across the network. In contrast, when switching to a
fixed \emph{total} BSM budget across the entire fabric, architectures with fewer
switches and shorter end-to-end paths benefit disproportionately: reduced hop
counts and lower accumulated optical loss allow switch-light designs to achieve
faster EPR generation despite a smaller number of fabric elements. Second, BCube exposes an additional coherence-driven
trade-off: increasing the communication-memory cutoff $\tau_{\mathrm{cut}}$
steadily improves BCube latency. Third, sensitivity to optical switch insertion loss is strongly topology dependent. In Clos and Fat-Tree, remote entanglement traverses many optical switches, causing loss to accumulate across hops and latency to increase rapidly as per-switch loss grows. In contrast, QFly and BCube are less sensitive due to shorter paths and, for BCube, the use of QPUs as repeaters that do not incur switch insertion loss in our model. Together, these results show that topology, resource placement,
and physical-layer parameters interact nontrivially, motivating architecture-aware
benchmarking for DQC.

\smallskip
The remainder of this paper is organized as follows.
Section$\S$~\ref{section:architcure} introduces the architectural characteristics and photonic components of each QDC topology.
Section$\S$~\ref{experiment} presents our simulation methodology and evaluation results.
Section~\ref{sec:conclusion} discusses design implications and future directions for co-optimizing topology, scheduling, and physical-layer control in scalable quantum data centers.

\section{Data Center Network Designs}
\label{section:architcure}

This section presents the architectural principles, parameterization, and
scaling behavior of the four quantum data-center architectures evaluated in this
work: Fat-Tree, Clos, QFly, and BCube. We focus on structural properties that
directly influence scalability, path diversity, switch count, and the role of
QPUs in communication, together with a high-level physical-layer loss and
entanglement-timing model. 

Distributed quantum computing performance depends on multiple factors, including circuit partitioning~\cite{kaur2025optimized,burt2024generalised,baker2020time}, compiler
transformations (e.g., circuit rewrites that adapt programs to architectural
constraints and reduce non-local communication) \cite{keskin2025lattice}, and entanglement-generation
and scheduling policies~\cite{monroe2014large,jiang2007distributed}. Modeling all such interactions jointly is
beyond the scope of this work. Instead, our objective is to establish a
controlled architectural baseline that isolates the impact of network topology
and shared photonic resources on distributed execution.

In this paper, we use the term fabric to denote the optical interconnection network—including switches, links, and Bell-state measurement resources that establishes and routes entanglement between QPUs.

\subsection{QDC Hardware Components and Entanglement Generation}
\label{sec:qdc_hw}

We summarize the core hardware components common to quantum data-center (QDC)
architectures and describe the entanglement-generation abstraction used
throughout this work.

Each QPU comprises a collection of \emph{data qubits}, which store and process
algorithmic quantum state, and a smaller number of \emph{communication qubits},
which interface with the photonic fabric to establish remote entanglement.
Communication qubits are required because entanglement is generated and heralded
outside the QPU — via photonic interference and detection — must be coherently
mapped onto local matter qubits before it can be consumed by computation.
Non-local two-qubit operations between QPUs are executed using
entanglement-assisted primitives, most commonly \emph{gate teleportation} or
\emph{state teleportation}, which consume pre-shared EPR pairs together with
local operations and classical communication. Compared to local gates, these
primitives incur additional latency and resource consumption, as they depend on
successful entanglement generation, Bell-state measurements, and timely
classical coordination. As a result, the availability and scheduling of
communication qubits, as well as contention for shared photonic resources,
directly shape the performance of distributed quantum execution.

A typical QDC optical fabric consists of the following building blocks:
(i) \emph{entangled photon-pair sources (EPPS)}, which probabilistically generate
pairs of entangled photons; (ii) \emph{optical switches (OSWs)}, which configure
photonic paths across racks, groups, or switching stages; (iii)
\emph{Bell-state measurement modules (BSMs)}, which perform two-photon interference and
herald successful entanglement generation or entanglement swapping; and (iv)
\emph{single-photon detectors (SPDs)}, often operated cryogenically to reduce
dark counts and timing jitter. Each QPU connects to the optical fabric through a
limited number of optical I/O ports; to increase concurrency, multiple
communication qubits may share a port via time-division multiplexing (TDM) or
wavelength-division multiplexing (WDM), at the cost of synchronization
and control overhead. In addition to these quantum components, QDC operation
relies on a classical control plane that provides clock distribution, phase
stabilization, heralding signals, and feedforward communication, which together
enable coordinated entanglement generation and distributed gate execution.

\paragraph{Entanglement-generation protocol.}
In this work, we focus on a \emph{scatter--scatter} (midpoint-interference)
entanglement-generation protocol, which is representative of near-term
photonic quantum networks and is widely adopted in experimental and architectural
studies. Alternative protocols—such as memory--memory or memory--source
links—are discussed extensively in the literature
(e.g.,~\cite{beukers2024remote}) and are left for future exploration.

\begin{figure}
    \centering
    \includegraphics[scale=0.43]{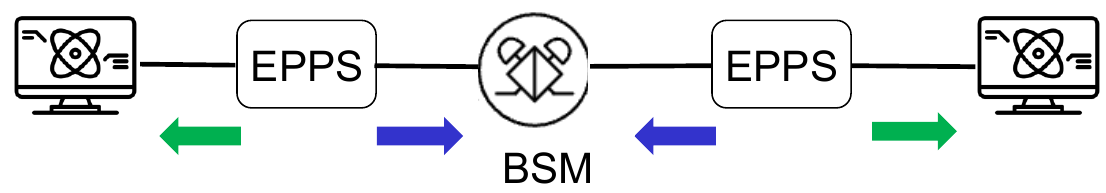}
    \caption{Scatter--scatter entanglement generation using independent (possibly non-degenerate) photon-pair sources and a central BSM.}
    \label{fig:scatter_scatter}
\end{figure}

As illustrated schematically in Fig.~\ref{fig:scatter_scatter}, the scatter--scatter
protocol employs two independent entangled photon-pair sources placed within the
optical fabric. During each entanglement attempt, one photon from each pair is routed
toward a central BSM, while the corresponding partner photons propagate toward
the two endpoint QPUs and interact with their respective communication qubits.
A successful end-to-end entanglement event is heralded by a coincidence detection
at the BSM together with the appropriate scattering or absorption events at the
two endpoint QPUs. Only upon these classical detection signals do the endpoint
communication qubits share a heralded Bell pair.

A key advantage of this protocol is its compatibility with \emph{non-degenerate}
entanglement sources, which emit photon pairs at different wavelengths (e.g.,
near-infrared and telecom). This enables efficient interfacing between QPUs and
low-loss optical fibers without relying on high-performance quantum frequency
conversion at the endpoints. As a result, scatter--scatter links are well suited
to heterogeneous photonic fabrics and near-term QDC deployments.

\subsection{Switch-Centric vs. Server-Centric}
\label{sec:switch_vs_server}

QDC architectures can be broadly classified as
\emph{switch-centric} or \emph{server-centric}, depending on where
entanglement-generation, switching, and swapping operations are performed.

\paragraph{Switch-centric architectures.}
In a switch-centric design, QPUs are connected by a switching fabric: they emit or absorb photons but do not store or
process intermediate entangled states.
End-to-end entanglement across multiple hops is realized through optical path
composition inside the fabric, typically requiring a single BSM operation per
Bell pair.
Fat-Tree, Clos, and QFly architectures fall into this category.

This design simplifies QPU requirements and
avoids requiring long-lived quantum memory at intermediate nodes.
However, it concentrates contention and loss within shared switching elements,
making performance sensitive to switch port budgets, BSM availability, and
reconfiguration delays.

\paragraph{Server-centric architectures.}
In server-centric designs, QPUs participate directly in 
entanglement distribution by acting as repeaters.
Intermediate QPUs store entangled states and perform Bell-state measurements to
extend entanglement over multiple hops.
BCube exemplifies this approach.

Compared to switch-centric architectures, server-centric designs trade reduced
dependence on complex optical switching hardware for increased reliance on QPU
capabilities.
Multi-hop entanglement requires multiple Bell-state measurements and
quantum memory hold times, making performance sensitive to communication-qubit
coherence and scheduling efficiency. 

This distinction has important system-level consequences. In a switch-centric
design, each multi-hop optical path requires only a single BSM operation for an
end-to-end Bell pair, whereas in a server-centric design, a repeater chain with
\(N_{\mathrm{rep}}\) intermediate QPUs requires \(N_{\mathrm{rep}}+1\) Bell-state
measurements and corresponding storage intervals. These additional operations
directly impact achievable entanglement rates, fidelity, and scheduling
complexity, motivating the architectural comparisons studied in the remainder
of this paper. This distinction parallels classical data-center networks, where switch-centric
Clos fabrics treat servers as pure endpoints, while server-centric designs such
as DCell and BCube involve servers directly in routing and forwarding
decisions~\cite{guo2008dcell,guo2009bcube}.
Analogously, in quantum data centers, switch-centric architectures centralize
entanglement operations in the fabric, whereas server-centric architectures
allow QPUs to participate in entanglement distribution.



\begin{table}[t]
\centering
\caption{Common notation used in the architecture overview.}
\label{tab:arch-notation}
\begin{tabular}{ll}
\toprule
Symbol & Meaning \\
\midrule
\(N\) & Number of QPUs (system size) \\
\(k\) & Port budget (ports per switch) \\
\(T\) & Number of ToR switches (Clos) \\
\(R\) & QPUs per rack (Clos) \\
\(n\) & BCube switch radix (\(n \triangleq k\)) \\
\(k_{\mathrm{bcube}}\) & BCube level parameter \\
\(k_{\text{ring}}\) & QFly inter-switch degree \\
\(N_{\text{rep}}\) & \# repeater QPUs on a BCube path \\
\bottomrule
\end{tabular}
\end{table}

 \paragraph{Notation.} 
Table~\ref{tab:arch-notation} summarizes the notation used throughout the paper.
We use \(N\) to denote the total number of QPUs in the
system. Each optical switch has a fixed \emph{port budget} \(k\), also referred
to as the \emph{switch radix}, which specifies the total number of input/output
ports available on the switch for connecting QPUs or other switches. For
Clos-style architectures, \(T\) denotes the number of top-of-rack (ToR) switches,
where a rack is a physical grouping of co-located QPUs and a ToR switch
aggregates and connects all QPUs within a single rack; \(R\) denotes the number
of QPUs attached to each rack.

For BCube, we consider a server-centric design in which QPUs participate in
routing; the BCube switch radix is set to \(n \triangleq k\), and
\(k_{\mathrm{bcube}}\) denotes the BCube \emph{level parameter}, i.e., the number
of switch layers in the BCube topology, which determines the network depth and
path diversity. For QFly, \(k_{\text{ring}}\) denotes the inter-switch degree,
i.e., the number of ports on each switch allocated to connections with other
switches, under the same total port budget \(k\). Finally, for server-centric
(BCube) routing, \(N_{\text{rep}}\) denotes the number of intermediate repeater
QPUs traversed along a communication path between two endpoint QPUs.

\subsection{Fat-Tree Architecture}

Fat-Tree is a canonical \emph{switch-centric} data-center network originally
proposed in the classical networking literature to provide full bisection
bandwidth (i.e., sufficient aggregate interconnect capacity to support
simultaneous full-rate communication between any equal partition of endpoints), predictable routing, and high path diversity at scale
\cite{leiserson1985fattree,alfares2008fattree}. Its design
philosophy is to realize high aggregate bandwidth using many moderate-radix
switches arranged in a symmetric multi-tier hierarchy, rather than relying on
expensive ultra-high-radix switches.

\begin{figure}
    \centering
    \includegraphics[scale=0.65]{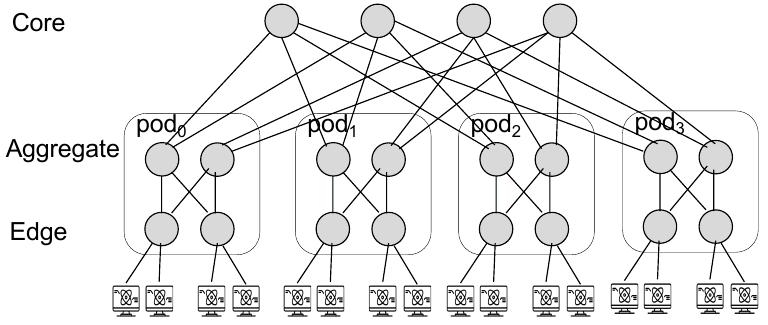}
    \caption{A Fat-Tree architecture with pods. Each of the \(k\) pods contains
\(k/2\) edge (top-of-rack) and \(k/2\) aggregation switches that connect QPUs to a
shared core layer of \((k/2)^2\) switches, providing full bisection bandwidth
under uniform switch radix \(k\).}

    \label{fig:fattree}
\end{figure}

The Fat-Tree fabric is organized into three logical layers as shown in figure \ref{fig:fattree}: edge (top-of-rack),
aggregation, and core. QPUs connect to top-of-rack switches . Edge switches aggregate intra-rack traffic,
aggregation switches provide connectivity among racks within a pod, and core
switches interconnect pods. QPUs attach to edge switches via dedicated access
ports. This regular hierarchy yields a symmetric topology with uniform path lengths and
balanced bandwidth, enabling predictable routing and efficient load balancing.

\paragraph{Parameterization and scaling.}
We characterize a Fat-Tree by its switch radix \(k\), representing the number of
ports per switch. We choose \(k\) such that it is  the smallest even
value \(k \ge 2\) for which the resulting topology can accommodate the target
system size \(N\):
\begin{equation}
k \;=\; \min\Bigl\{k' \in 2\mathbb{Z}_{\ge 2} \;:\; \tfrac{{k'}^{3}}{4} \ge N \Bigr\}.
\end{equation}

For a given \(k\), the network comprises \(k\) pods, where each pod is a modular
subnetwork containing \(k/2\) edge (top-of-rack) switches, \(k/2\) aggregation
switches, and the QPUs attached to those edge switches. The network further
includes a core layer of \((k/2)^2 = k^{2}/4\) switches: 
\begin{equation}
S_{\text{edge}}=\tfrac{k^{2}}{2},\quad
S_{\text{agg}}=\tfrac{k^{2}}{2},\quad
S_{\text{core}}=\tfrac{k^{2}}{4},\quad
S_{\text{total}}=\tfrac{5k^{2}}{4}.
\end{equation}
Each edge switch serves up to \(k/2\) QPUs, which sets the maximum rack capacity
to
\begin{equation}
H_{\text{per-edge}}=\tfrac{k}{2}.
\end{equation}

This construction follows the canonical \(k\)-ary Fat-Tree topology of
Al-Fares \emph{et al.}~\cite{al2008scalable}.



\subsection{Clos Architecture}

Clos is a flexible \emph{switch-centric} architecture that generalizes the
Fat-Tree design by decoupling system scale from a single global radix parameter.
While the canonical data-center Fat-Tree corresponds to a specific symmetric
instantiation of a folded-Clos network, the broader Clos framework admits a wide
range of valid realizations obtained by independently dimensioning switch radix,
rack fanout, and fabric size.
As in Fat-Tree, QPUs act strictly as endpoints and connect to a multi-stage
optical switching fabric.

\paragraph{High-level organization.}
Clos networks employ a multi-stage switching fabric composed of top-of-rack
(ToR), aggregation, and core layers.
QPUs attach to ToR switches via dedicated access ports, while the remaining ports
are used for inter-switch connectivity across stages.
Unlike Fat-Tree, which enforces a fixed and symmetric relationship between the
number of switches and port allocations, Clos allows these parameters to vary,
enabling finer-grained control over rack density and overall fabric scale.

\paragraph{Design flexibility.}
A Clos fabric can be instantiated in infinitely many ways depending on how
ToR fanout and switch radix are chosen.
Given \(T\) ToR switches and \(R\) QPUs per ToR, the fabric supports up to
\(T \times R\) QPUs.
For a target system size \(N\), any configuration satisfying
\(T \times R \ge N\) is feasible, with the difference \(T \times R - N\)
corresponding to \emph{unused rack capacity}, i.e., unoccupied host attachment
slots arising from rack-level granularity.

Different Clos designs reflect different architectural priorities.
A \emph{Clos\_tight} design emphasizes dense rack packing by selecting parameters
that keep unused rack capacity minimal, resulting in closely matched fabric and
system sizes.
In contrast, a \emph{Clos\_compact} design emphasizes reducing the total number
of switches in the fabric, allowing modest unused rack capacity when doing so
simplifies the switching infrastructure.
Both represent valid points within the broader Clos design space and share the
same fundamental wiring principles.

\paragraph{Implications.}
Compared to Fat-Tree, Clos provides finer-grained control over system scaling and
rack density without requiring a single global radix to determine capacity.
This flexibility enables tighter matching between system size and physical
resources, but introduces additional degrees of freedom that can increase switch
count and coordination overhead depending on how the fabric is dimensioned.

\subsection{QFly Architecture}

QFly is a flattened, low-diameter \emph{switch-centric} architecture inspired by
Dragonfly-style interconnects, which employ groups of high-radix routers and
sparse global links to achieve low diameter while limiting long-cable
cost~\cite{kim2008technology}. In QFly, QPUs are grouped under a smaller number of
high-radix optical switches, reducing hierarchy depth and the number of fabric
elements compared to Fat-Tree and Clos.

\paragraph{Implications.}
Increasing inter-switch connectivity reduces network diameter and increases path
diversity, but concentrates traffic onto fewer high-radix switching elements.

\begin{figure}
    \centering
    \includegraphics[scale=0.5]{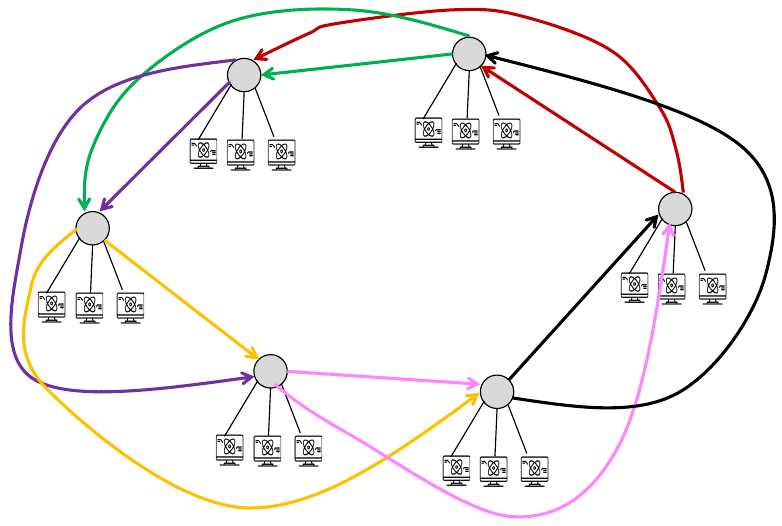}
\caption{Example QFly topology with \(S=6\) switches, each serving \(m=3\) QPUs (\(N=18\)). Each switch uses \(k_{\mathrm{ring}}=2\) ports for inter-switch connectivity. By increasing \(k_{\mathrm{ring}} \) we can reduce the number of hops between QPUs.
}
    \label{fig:qfly}
\end{figure}

\subsection{BCube Architecture}

BCube is a \emph{server-centric} architecture in which QPUs connect to multiple
switch layers and can participate directly in entanglement distribution. Unlike
switch-centric designs, BCube can form end-to-end entanglement via \emph{repeater
chains}: intermediate QPUs generate, store, and swap EPR pairs on behalf of other
QPUs, so QPUs serve dual roles as compute nodes and repeaters.

\begin{figure}
    \centering
    \includegraphics[scale=0.65]{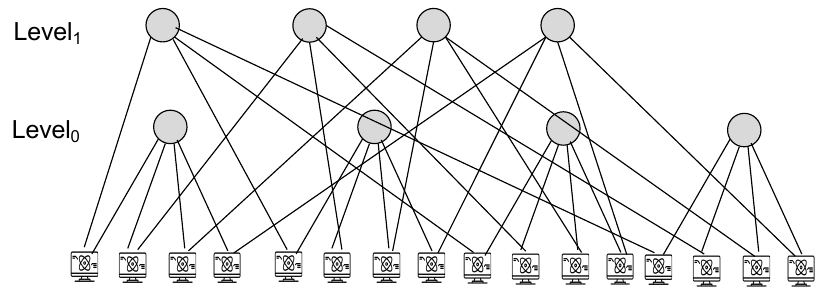}
\caption{Example \((4,2)\)-BCube topology with 16 QPUs. QPUs connect to two switch
layers, with four switches per level. Communication paths may traverse
intermediate QPUs, which act as repeaters in the server-centric BCube design.}

    \label{fig:bcube}
\end{figure}

We follow the original BCube construction of Guo \emph{et al.}~\cite{guo2009bcube}.
An \((n,k_{\mathrm{bcube}})\)-BCube consists of \(k_{\mathrm{bcube}}+1\) switch
layers, each containing \(n^{k_{\mathrm{bcube}}}\) switches of radix \(n\).
Each QPU connects to exactly one switch in every layer, and the total system
capacity scales as \(n^{k_{\mathrm{bcube}}+1}\) QPUs as shown in figure \ref{fig:bcube}.

QPUs are indexed by \((k_{\mathrm{bcube}}+1)\)-digit identifiers in base \(n\),
and switch connectivity is defined by fixing one digit position per layer.
As a result, when the number of instantiated QPUs is smaller than the full
\(n^{k_{\mathrm{bcube}}+1}\) capacity, only the subset of switches whose indices
are referenced by at least one QPU are \emph{active}, while the remaining switches
are structurally present but unused. This distinction is important for accurately
accounting for switch count, port utilization, and available Bell-state
measurement (BSM) resources in partially populated BCube instances.

\paragraph{Repeater-chain implications.}
For a path with \(N_{\text{rep}}\) intermediate repeater QPUs, establishing an
end-to-end Bell pair requires generating entanglement on \(N_{\text{rep}}+1\)
segments and performing \(N_{\text{rep}}+1\) Bell-state measurements (one per
swap).


\subsection{Optical Loss and EPR-Generation Timing Model}
\label{sec:loss-model}

We model the time required to generate entanglement in different architecures for non-local gates. A \emph{non-local two-qubit gate} is a gate whose two operand qubits
reside on different QPUs. To execute such a gate, a scheduler selects a
\emph{path} between the two endpoint QPUs and requests the
generation of an Einstein--Podolsky--Rosen (EPR) pair along that path. For each
non-local gate, we estimate the \emph{expected} EPR-generation latency
$\mathbb{E}[T_{\mathrm{pair}}]$ associated with the scheduler-selected path.

\paragraph{paths and hop count.}

A path is an ordered sequence of physical links connecting the two endpoint QPUs
that host the operand qubits of a non-local gate. These links consist of
QPU--switch, switch--switch, and switch--QPU connections, depending on the
architecture.

In switch-centric architectures (QFly, Clos, and Fat-Tree), QPUs act strictly as
endpoints. A path starts with a QPU--switch link from the source QPU, traverses
one or more switch--switch links within the optical switching network, and ends
with a switch--QPU link to the destination QPU. Intermediate QPUs do not
participate in entanglement distribution. Regardless of path length, a multi-hop
path requires only a single Bell-state measurement (BSM) within the switching
network to establish end-to-end entanglement.

In server-centric architectures (BCube), QPUs may act as repeaters. A path
consists of a sequence of QPU--switch--QPU segments, where each intermediate QPU
stores entanglement and performs entanglement swapping to extend the connection.
As a result, entanglement is generated hop by hop: each adjacent QPU pair
requires BSM capacity at the connecting switch, and each intermediate QPU must
have available communication qubits.


\paragraph{Per-attempt timing.}
Entanglement is generated probabilistically through repeated attempts. Let
$T_{\mathrm{att}}$ denote the duration of a single entanglement-generation
attempt on a given path. We model
\begin{equation}
T_{\mathrm{att}}
=
T_{\mathrm{src}}
+
\frac{2D}{v_{\mathrm{fiber}}}
+
T_{\mathrm{reset}},
\end{equation}
where $T_{\mathrm{src}}$ is the source repetition time,
$T_{\mathrm{reset}}$ is the communication-qubit reset or reinitialization time,
$v_{\mathrm{fiber}}$ is the speed of light in fiber, and $D$ is the total fiber
length traversed by photons along the selected path.

\paragraph{Optical loss model.}
For a given path, the total optical loss (in dB) is modeled as
\begin{equation}
L_{\mathrm{tot}}
=
L_{\mathrm{fiber}}
+
L_{\mathrm{sw}}
+
L_{\mathrm{BSM}}
\; (+\, L_{\mathrm{mem}}),
\label{eq:loss_total}
\end{equation}
where $L_{\mathrm{fiber}}$ captures fiber attenuation,
$L_{\mathrm{sw}}$ is the cumulative insertion loss from optical switches,
and $L_{\mathrm{BSM}}$ abstracts inefficiencies in Bell-state measurement
operations (e.g., coupling and detector loss). The term $L_{\mathrm{mem}}$
applies only to server-centric paths and captures additional loss due to storing
entanglement in intermediate QPU memories, including photon--matter conversion
loss and decoherence during storage.

\paragraph{Switch insertion loss.}
Photons traversing an optical switch incur insertion loss due to the internal
multi-stage switching structure. With switch radix $k$ and per-$2\times2$
element loss $\ell_{2\times2}$ (dB), the loss of a single switch traversal is
approximated as
\[
L_{\mathrm{sw}}(k)
\approx
(2\lceil \log_2 k\rceil - 1)\,\ell_{2\times2},
\]
corresponding to a Bene\v{s}-style non-blocking optical switch design
\cite{spanke1987n}. Let $n_{\mathrm{sw}}$ denote the number of switch traversals
along the path; then $L_{\mathrm{sw}} = n_{\mathrm{sw}}\,L_{\mathrm{sw}}(k)$.

\paragraph{Expected EPR-generation latency.}
Total loss is converted to an end-to-end transmittance
$T_{\mathrm{chan}} = 10^{-L_{\mathrm{tot}}/10}$, which we interpret as the
per-attempt success probability for generating an EPR pair along the selected
path. For switch-centric architectures, assuming independent attempts, the
expected EPR-generation latency is
\begin{equation}
\mathbb{E}[T_{\mathrm{pair}}]
\approx
\frac{T_{\mathrm{att}}}{T_{\mathrm{chan}}}.
\label{eq:epr_dt_switch}
\end{equation}

\paragraph{Server-centric retry window and protocols.}
In server-centric architectures such as BCube, entanglement must be generated on
multiple segments and combined through entanglement swapping at intermediate
QPUs. Because partially generated entanglement must be stored while waiting for
other segments to succeed, we impose a coherence-limited \emph{retry window}
$\tau_{\mathrm{cut}}$. If an end-to-end EPR pair has not been successfully formed
via swapping within $\tau_{\mathrm{cut}}$, the attempt is discarded and restarted.
We evaluate two repeater-chain scheduling protocols—\emph{sequential} and
\emph{parallel}—introduced in~\cite{pouryousef2024minimal}, which capture
different trade-offs between control simplicity and latency. For these
protocols, we estimate the raw end-to-end entanglement delivery rate
$R_{\mathrm{raw}}$ and compute
\begin{equation}
\mathbb{E}[T_{\mathrm{pair}}]
\approx
\frac{1}{R_{\mathrm{raw}}}.
\label{eq:epr_dt_mem}
\end{equation}

\begin{table}[t]
\centering
\caption{Notation for the optical loss and timing model.}
\label{tab:loss-model-notation}
\begin{tabular}{ll}
\toprule
Symbol & Description \\
\midrule
$h$ & Number of hops in a path \\
$\ell$ & Fiber length per segment (km) \\
$\alpha$ & Fiber attenuation (dB/km) \\
$k$ & Switch radix \\
$\ell_{2\times2}$ & Loss of one $2\times2$ switch element (dB) \\
$L_{\mathrm{fiber}}$ & Total fiber loss (dB) \\
$L_{\mathrm{sw}}$ & Total switch insertion loss (dB) \\
$L_{\mathrm{BSM}}$ & Bell-state measurement loss (dB) \\
$L_{\mathrm{mem}}$ & Per-hop memory loss (dB) \\
$T_{\mathrm{chan}}$ & End-to-end transmittance \\
$T_{\mathrm{coh}}$ & Communication-qubit coherence cutoff \\
$\mathbb{E}[T_{\mathrm{pair}}]$ & Expected EPR-generation latency \\
\bottomrule
\end{tabular}
\end{table}

\section{Evaluation}
\label{experiment}

In this section, we evaluate multiple variants of the QFly, BCube, Fat-Tree, and
Clos architectures using three targeted experiments. Each experiment isolates a
specific physical or architectural factor that influences the execution latency
of distributed quantum circuits.

\paragraph{Workloads and system scale.}
A \emph{workload} is a family of circuits specified by (i) the total logical width
(number of data qubits) and (ii) a two-qubit \emph{interaction pattern} that determines which qubit pairs are considered for two-qubit gates. We consider three workload families commonly used in quantum benchmarking:
(i) \emph{Nearest-neighbor circuits}, where all two-qubit gates act on adjacent
qubits in a 1D chain, representing highly local, hardware-friendly patterns;
(ii) \emph{Random Clifford+T circuits}, consisting of mixed single-qubit
Cliffords, scattered CNOTs, and sparse $T/T^\dagger$ gates, approximating
generic fault-tolerant workloads with moderate non-locality; and
(iii) \emph{Long-range circuits}, where two-qubit gates couple uniformly random
pairs of qubits, producing communication-heavy programs with high non-locality.

System size is the number of QPUs, denoted by $N$. The total circuit
width is set proportional to system scale by assigning each QPU a fixed local
grid of data qubits. We evaluate multiple \emph{system scales}
by varying $N$ while keeping the per-QPU data-qubit capacity and the number of
communication qubits per QPU fixed. Thus, increasing $N$ increases the \emph{total}
circuit width proportionally (more QPUs, each with the same local qubit grid). The resulting circuits are then partitioned using the same
compilation and mapping pipeline and simulated under architecture-specific
routing and scheduling policies (we explain this later). Circuits are partitioned using a
lightweight Kernighan--Lin (KL) heuristic \cite{KL_algorithm}. For the \emph{monolithic baseline}, we assume a single QPU that hosts all physical
qubits required to handle the circuit, supports full all-to-all connectivity, and incurs no inter-QPU
communication or entanglement-generation overhead. The circuit execution latency
in this case is denoted by $T_{\mathrm{mono}}$.

\paragraph{Key physical parameters.} Here we define the key parameters that we change in the system. \emph{Switch insertion loss} is the optical attenuation (in dB) incurred when a photon
traverses a switching element. \emph{Switch reconfiguration delay} is the time required to reconfigure an optical switch
before it can support a new set of connections (e.g., setting up new optical paths in the network).
\emph{memory cutoff or memory coherence time} (or \emph{coherence cutoff}) is denoted by $\tau_{\mathrm{cut}}$
and is the maximum time an entangled qubit may be stored at a QPU
(e.g., at a repeater) before it is discarded. This parameter is
only relevant for server-centric architectures (e.g., \textsc{BCube}).

\paragraph{Execution model and schedulers.}
We represent each quantum circuit as a directed acyclic graph (DAG), where nodes
correspond to gates and edges encode dependencies between gates. At any
time, there is a set of \emph{ready} gates (called frontier layer) whose predecessors have all
completed. We remove completed gates from the DAG, which may expose new ready
gates and form the next frontier. For each DAG layer (frontier), we process the ready non-local gates in a random
order. For each gate, the scheduler attempts to reserve the required
communication resources by invoking the path-selection procedure: it searches for a feasible qubit-to-qubit path. If a feasible path is found, the corresponding
resources are reserved immediately; otherwise, the gate is deferred and remains
in the frontier for the next scheduling step. We advance simulated time by the latency of the
\emph{slowest} successfully established remote entanglement in that layer
(including switch reconfiguration delay), commit the completed gates,
release their resources, and repeat the procedure for the next layer. After parsing and serving all the gates in the DAG, we report the final value of the time variable as circuit execution latency.

\paragraph{Path selection and scheduling.}
For each non-local two-qubit gate, the scheduler searches for a
\emph{shortest feasible qubit-to-qubit path} between the two QPUs that host
the gate’s operand qubits. Candidate paths are enumerated in increasing hop count
order and tested sequentially. A path is feasible only if all required resources are available:
sufficient capacity on each fabric edge (optical channels), sufficient communication
qubits at the endpoint QPUs, and sufficient BSM
resources. In switch-centric architectures (QFly, Clos, Fat-Tree), a multi-hop path requires
only one BSM at the fabric to establish end-to-end entanglement. In BCube,
entanglement is generated hop by hop: each intermediate QPU must have available
communication qubits to act as a repeater, and each adjacent QPU pair requires
BSM capacity at the connecting switch.

Among shortest feasible paths, the scheduler greedily selects the first
valid candidate and immediately reserves the required resources. Circuit execution is driven by an event-based scheduler that uses
expected EPR-generation latencies derived from stochastic loss and accounts for contention for BSMs and switch reconfiguration
delays.

For a given distributed architecture, we denote the corresponding end-to-end
execution latency by $T_{\mathrm{dist}}$. We report results using the
\emph{distributed-to-monolithic latency ratio}
\begin{equation}
\rho_{\mathrm{lat}} \;\triangleq\; \frac{T_{\mathrm{dist}}}{T_{\mathrm{mono}}},
\end{equation}
which isolates the performance overhead introduced purely by distribution and
communication. Values of $\rho_{\mathrm{lat}}$ close to one indicate near-ideal
scaling relative to the monolithic baseline, while larger values reflect
increasing communication and scheduling overheads.


\subsection{Sanity Checks: Remote-Gate Timing Under Varying Parameters}
\label{sec:model-validation}

We begin by performing sanity checks to ensure that the
simulator responds sensibly to basic physical trends. These experiments are not intended to validate the 
model but rather to verify that loss, retry, and storage effects behave as
expected.

We consider paths with $h=4$ hops, where adjacent components (QPU--switch or switch--switch) are separated by $0.1$\,km,
with fiber attenuation of $0.2$\,dB/km. Communication-qubit memory loss due to
storage and swapping in the server-centric case is fixed at $3$\,dB per QPU. The
EPR source rate is $10^{6}$ pairs/s, and the local two-qubit gate time is
$T_{\mathrm{local}}=10\,\mu$s. Switch reconfiguration delay is excluded to isolate
loss and memory effects.

For the \texttt{Server}-centric paradigm, we evaluate both \textbf{sequential}
(\texttt{S}) and \textbf{parallel} (\texttt{P}) swap-scheduling protocols proposed
in~\cite{pouryousef2024minimal}, with a memory cutoff time $\tau_{\mathrm{cut}}$
that limits how long a partially generated entangled state may be stored if the
required swaps have not yet been completed, after which the attempt is discarded. We report two derived, dimensionless metrics to summarize the behavior of the
optical loss and entanglement-timing model.

First, we define the \emph{EPR-generation latency ratio} as
\begin{equation}
\rho_{\mathrm{EPR}} \;\triangleq\;
\frac{\mathbb{E}[T_{\mathrm{pair}}]}{T_{\mathrm{local}}},
\end{equation}
where $\mathbb{E}[T_{\mathrm{pair}}]$ is the expected latency to generate the
end-to-end EPR pair required for executing a non-local two-qubit gate under the
selected route and entanglement-generation protocol (cf.\
Eqs.~\eqref{eq:epr_dt_switch}--\eqref{eq:epr_dt_mem}), and
$T_{\mathrm{local}}$ is the duration of a local two-qubit gate. This ratio
captures the relative cost of remote entanglement generation compared to local
computation.

Second, we define the \emph{switch-to-memory loss ratio} as
\begin{equation}
\rho_{\ell} \;\triangleq\;
\frac{L_{\mathrm{sw}}^{\mathrm{(hop)}}}{L_{\mathrm{mem}}},
\end{equation}
where $L_{\mathrm{sw}}^{\mathrm{(hop)}}$ denotes the insertion loss incurred by a
single switch traversal. Following a Bene\v{s}-style optical switch fabric, we
approximate
$L_{\mathrm{sw}}^{\mathrm{(hop)}} = (2\lceil \log_2 k\rceil - 1)\,\ell_{2\times2}$,
where $k$ is the switch radix and $\ell_{2\times2}$ is the loss of a single
$2\times2$ switching element. In this experiment, we fix the switch radix to
$k=2$, yielding $L_{\mathrm{sw}}^{\mathrm{(hop)}}=\ell_{2\times2}$. The term
$L_{\mathrm{mem}}$ is the per-hop memory-induced loss incurred when a QPU stores
and swaps entanglement while acting as an intermediate repeater. This ratio
characterizes the relative penalty of switch-centric versus server-centric
entanglement distribution as path length increases.


\begin{figure} \centering \includegraphics[scale=0.45]{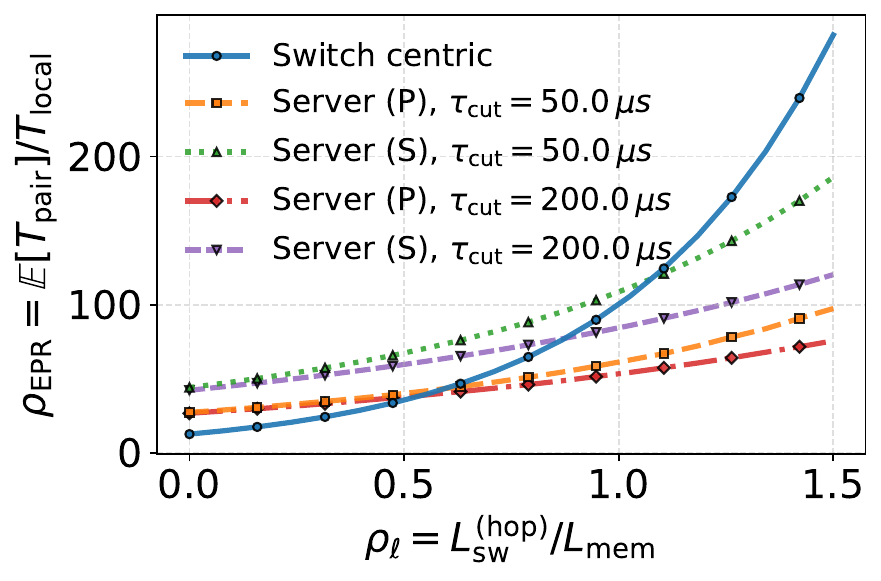} \caption{EPR-generation latency ratio
$\rho_{\mathrm{EPR}}=\mathbb{E}[T_{\mathrm{pair}}]/T_{\mathrm{local}}$
as a function of the per-hop switch-to-memory loss ratio
$\rho_{\ell}=L_{\mathrm{sw}}^{(\mathrm{hop})}/L_{\mathrm{mem}}$
for a four-hop path. We compare switch-centric entanglement distribution with
server-centric (QPU-assisted) sequential and parallel protocols under two
communication-qubit coherence cutoffs. Server-centric paths incur fewer switch
traversals and benefit from parallel EPR generation when coherence times are
long, while switch-centric paths perform better when memory-induced loss
dominates.}
\label{fig:server_switch_centric_ratio_multi_hops} 
\end{figure}

Figure~\ref{fig:server_switch_centric_ratio_multi_hops} reports the
\emph{EPR-generation latency ratio}
$\rho_{\mathrm{EPR}}=\mathbb{E}[T_{\mathrm{pair}}]/T_{\mathrm{local}}$
for a fixed multi-hop path as a function of the
\emph{switch-to-memory loss ratio}
$\rho_{\ell}=L_{\mathrm{sw}}^{\mathrm{(hop)}}/L_{\mathrm{mem}}$.
In switch-centric architectures, photons traverse only the optical fabric,
incurring switch insertion loss at every hop but avoiding memory-induced loss.
In contrast, server-centric architectures incur fewer switch traversals but rely
on intermediate QPUs to store and swap entanglement, making them sensitive to
memory loss and coherence constraints.

When the communication-qubit coherence cutoff $\tau_{\mathrm{cut}}$ is small,
server-centric paths experience frequent memory expiration events, inflating
$\mathbb{E}[T_{\mathrm{pair}}]$ and degrading performance as explained in \cite{pouryousef2024minimal}. As
$\tau_{\mathrm{cut}}$ increases (e.g., from $50\,\mu$s to $200\,\mu$s), memory
expirations are reduced and \emph{parallel} entanglement-generation protocols
enable concurrent link establishment across hops, significantly lowering the
expected EPR-generation latency relative to \emph{sequential} protocols.

Conversely, when $\rho_{\ell}$ approaches zero—i.e., when memory-induced loss
dominates over switch insertion loss—switch-centric architectures outperform
server-centric ones regardless of protocol choice, since server-centric paths
incur an unavoidable memory-loss penalty from storing and retrieving
entangled qubits.


\subsection{Workload and Scale Sensitivity}
\label{sec:workload-scale}

This experiment evaluates how circuit execution latency varies across both
workload structure and system scale for four representative quantum
data-center architectures (QFly, BCube, and Fat-Tree, and Clos). We fix the underlying
physical-layer parameters—including fiber attenuation, switch insertion loss,
switch reconfiguration delay, and per-QPU data-qubit capacity—and examine two
axes of variation: (i) the \emph{type of circuit workload} and (ii) the
\emph{number of QPUs} in the system.

\subsection{Architecture Instantiations}

To enable a fair and reproducible comparison, we first describe how each quantum
data-center architecture is instantiated for a given system size \(N\). For each topology, we follow standard constructions from the literature and
widely adopted best practices, selecting parameters that respect realistic
switch port budgets, scale to the target number of QPUs, and align
physical-layer assumptions across architectures as closely as possible.
Where exact alignment is not feasible, we preserve architectural feasibility;
for example, in BCube, forcing QPUs to have the same port count as in Fat-Tree
can render the topology disconnected even with an arbitrarily large number of
switches, and is therefore not imposed.

In our model, switch ports serve three distinct roles: (i) connecting switches to
other switches, (ii) connecting switches to QPUs, and (iii) interfacing with BSM modules (Bell state analyzers and photon detectors) to perform entanglement swapping. To avoid
conflating architectural topology with entanglement-generation resources such as BSAs and photon detectors, we
first construct the logical network topology of each architecture using switch
ports only for inter-switch and switch--QPU connectivity. BSM resources are incorporated \emph{after} the topology is instantiated,
allowing us to isolate the effects of classical network structure from the
provisioning of quantum entanglement-generation resources.

We consider two complementary models for allocating BSM resources to switches.
In the first model, we assume a \emph{fixed number of BSMs per switch}. Under this
assumption, once the topology of each architecture is instantiated, each switch
is provisioned with additional ports dedicated to BSM modules, causing the total
BSM budget to scale with the number of switches. This model isolates the impact
of architectural structure when entanglement-generation capability is locally
provisioned rather than globally constrained.

In the second model, we assume a \emph{fixed total BSM budget} shared across the
entire architecture. This budget is distributed evenly across all switches,
implying that architectures with different numbers of switches receive different
per-switch BSM allocations. As a result, architectures with fewer switches receive a higher number of BSMs per switch and therefore require more BSM-facing ports on each switch, while deeper or switch-rich
architectures divide the same budget across a larger number of switching
elements.


\paragraph{Fat-Tree and Clos.}
For Fat-Tree, we instantiate the topology using the parameterization
and scaling equations introduced in Section~\ref{section:architcure}, selecting the
minimum feasible switch radix that supports the target system size \(N\). The
resulting Fat-Tree switch radix is then used as a common reference when
parameterizing the QFly and BCube architectures, ensuring consistent port budgets
across all evaluated topologies.

\paragraph{Clos instantiation.}
Following the design space described in Section~\ref{section:architcure},
we instantiate Clos topologies using an automated layout procedure that derives
a feasible configuration for each target system size \(N\).
The procedure selects an even switch radix \(k\), determines the number of
top-of-rack (ToR) switches \(T = k^{2}/4\), and assigns the number of QPUs per
rack \(R\) such that \(T \times R \ge N\), while respecting practical constraints
on rack fanout and ToR port budgets.
This construction follows the standard folded-Clos model, where each ToR switch
allocates \(k/2\) ports to QPUs and \(k/2\) ports to uplinks
\cite{clos1953study,greenberg2009vl2,singh2015jupiter}.

To reflect different points in the Clos design space, we use two instantiation
policies.
\emph{Clos\_tight} prioritizes minimizing unused rack capacity
(\(T \times R - N\)), yielding densely packed racks.
\emph{Clos\_compact} prioritizes minimizing the total number of switches,
allowing limited unused rack capacity when this significantly reduces fabric
size.
For each policy, once \((k, T, R)\) is determined, the aggregation and core
layers, inter-stage connectivity, and path structure are derived
deterministically.

\paragraph{Qfly}
QFly is instantiated under the same per-switch port budget \(k\) as the Fat-Tree
baseline. Each switch devotes \(m = k/2\) ports to attached QPUs, matching the
Fat-Tree edge fanout.

Let \(S\) denote the number of QFly switches and define
\(k_{\text{ring}}\) as the number of inter-switch ports per switch. We consider
three QFly variants that differ in how inter-switch connectivity is realized:
\begin{itemize}
\item \emph{QFly (\(k_{\mathrm{ring}} = m/2\)):}
With \(m = k/2\) ports allocated to QPUs, only \(m/2\) ports per switch are used
for inter-switch connectivity, yielding partial connectivity among switches.

\item \emph{QFly (\(k_{\mathrm{ring}} = k - m\)):}
With \(m = k/2\) ports reserved for QPU attachment, all remaining \(k - m\)
ports per switch are used for inter-switch connectivity, maximizing switch-level
path diversity under a fixed port budget.

  \item \emph{QFly (\(k_{\mathrm{ring}} = N/m - 1\)):}
  Switches are assumed to be fully connected, corresponding to an idealized
  upper bound that exceeds the original port budget.
\end{itemize}

\paragraph{BCube}:
We instantiate BCube using the same port budget as Fat-Tree by setting
\(n \triangleq k\). We choose the smallest \(k_{\mathrm{bcube}}\) such that:
\begin{equation}
k_{\mathrm{bcube}} = \left\lceil \log_{n}(N) \right\rceil - 1.
\end{equation}
BCube has \(L = k_{\mathrm{bcube}} + 1\) switch levels; each level contains
\(n^{k_{\mathrm{bcube}}}\) switches:
\begin{equation}
S_{\text{per-level}} = n^{k_{\mathrm{bcube}}}, \qquad
S_{\text{total}} = (k_{\mathrm{bcube}}+1)\,n^{k_{\mathrm{bcube}}}.
\end{equation}

Table \ref{table:QPU_scale_switches_updated} summarizes the derived architectural parameters used in the
experiments. For each architecture and QPU count, the table reports the total
number of switches, the number of QPUs per rack (or edge switch), and the number
of available BSMs per switch. These parameters are determined by the topology
construction rules of each architecture (e.g., Clos layers, BCube levels, or
QFly rings) and are not independently tuned. As a result, the table reflects
the inherent scaling properties of each architecture rather than additional
optimization.

As an example, consider BCube scaling from $N=64$ to $N=128$ QPUs.
We derive the per-switch port budget $k_{\mathrm{FT}}$ from the Fat-Tree baseline
as $k_{\mathrm{FT}}=\lceil(4N)^{1/3}\rceil$, rounded to the next even integer,
which yields $k_{\mathrm{FT}}=8$.
In the $(n,k)$-BCube construction, we set $n=k_{\mathrm{FT}}$ and choose
$k=\lceil\log_n N\rceil-1$, giving $k=1$ for $N=64$ and $k=2$ for $N=128$.
For $N=64$, this yields two switch levels, and since QPU identifiers occupy only a
single value of the most significant base-$n$ digit, only $16$ switches are
active.
For $N=128$, the additional level activates a third layer of switches; although
the total fabric contains $(k+1)n^k=192$ switches, only those connected to at
least one QPU are active, resulting in $16$, $16$, and $64$ active switches
across the three levels, for a total of $96$.
The jump in switch count from $16$ to $96$ therefore arises from a configuration
boundary in BCube, where increasing $N$ forces the introduction of an additional
level in the multi-stage fabric.

\begin{figure*}
    \centering
    \includegraphics[scale=0.73]{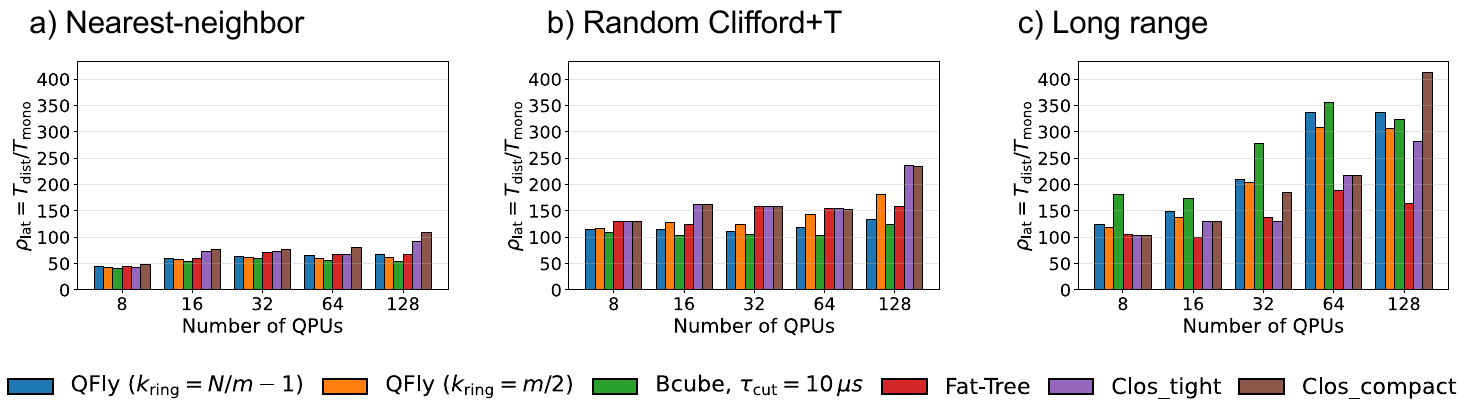}
    \caption{Latency ratio (Distributed / Monolithic) versus number of QPUs for each workload under the
\emph{fixed BSMs per switch} resource model. Bars show the average ratio across repeated runs for all architectures present in the dataset. Lower values indicate better scalability relative to the monolithic baseline.}
    \label{fig:workload_scalability}
\end{figure*}

For Fig.~\ref{fig:workload_scalability}, \emph{fixed BSMs-per-switch} model, where each switch is provisioned with exactly
two BSMs, causing the total BSM budget to scale with the number of switches. In contrast, Fig.~\ref{fig:scale_BSM_per_swtich} adopts a fixed global budget of BSMs that is evenly distributed
across all switches in each architecture.
All fixed parameters for this experiment are summarized in
Table~\ref{tab:workload_scale_params}. Unlike approaches that restrict BSM placement to specific layers, we distribute
BSM resources across \emph{all} switch layers in the instantiated topology.

\begin{table}[t]
\centering
\caption{Fixed physical and system parameters used in the workload scalability
experiment.}
\label{tab:workload_scale_params}
\begin{tabular}{lc}
\toprule
Parameter & Value \\
\midrule
QPU capacity (data qubits per QPU) & 16 \\
Communication qubits per QPU & 5 \\
Optical channels per fiber & 5 \\
QPU--switch distance & 0.1 km \\
Switch--switch distance & 0.1 km \\
Fiber attenuation & 0.1 dB/km \\
Communication-qubit memory loss & 3 dB \\
Switch insertion loss & 0.3 dB \\
Communication-qubit reset time & $T_{\mathrm{reset}} = 10\,\mu$s \\
EPR source generation rate & $10^{6}$ pairs/s \\
Coherence cutoff (BCube) & $\tau_{\mathrm{cut}} = 200\,\mu$s \\
Total BSM budget & 100 \\
\bottomrule
\end{tabular}
\end{table}

\paragraph{Scaling the number of QPUs.}
To assess scalability, we repeat the experiment for increasing system sizes
(e.g., $N \in \{8,16,32,64,128\}$ QPUs). For each architecture and QPU count, we
generate a circuit from the selected workload family with the number of two-qubit
gates scaled with $N$ (ranging from 200 at $N{=}8$ to 600 at $N{=}128$), partition
it across the available QPUs using the same compilation pipeline, and simulate
its distributed execution. We report scalability using the
distributed-to-monolithic latency ratio $\rho_{\mathrm{lat}}$, which captures
how execution latency grows with system size relative to the monolithic
baseline.


\begin{table}[t]
\centering
\small
\resizebox{\columnwidth}{!}{%
\begin{tabular}{lrrrrr}
\toprule
Arch. & \#QPUs & \#Sw. & QPUs/Rack & Ports/Sw. & \#ToRs \\
\midrule
QFly ($k_{\mathrm{ring}} = N/m - 1$) & 16  & 8  & 2.00 & 9/9   & 8  \\
QFly ($k_{\mathrm{ring}} = m/2$)     & 16  & 8  & 2.00 & 6/6   & 8  \\
QFly ($k_{\mathrm{ring}} = k - m$)   & 16  & 8  & 2.00 & 4/4   & 8  \\
Clos\_tight                          & 16  & 28 & 1.00 & 8.00  & 16 \\
Clos\_compact                        & 18  & 18 & 2.00 & 6.00  & 9  \\
Fat-Tree                             & 16  & 20 & 2.00 & 4.00  & 8  \\
BCube                                & 16  & 8  & 2.00 & 4.00  & 8  \\
\midrule
QFly ($k_{\mathrm{ring}} = N/m - 1$) & 64  & 16 & 4.00 & 19/19 & 16 \\
QFly ($k_{\mathrm{ring}} = m/2$)     & 64  & 16 & 4.00 & 12/12 & 16 \\
QFly ($k_{\mathrm{ring}} = k - m$)   & 64  & 16 & 4.00 & 8/8   & 16 \\
Clos\_tight                          & 64  & 28 & 4.00 & 8.00  & 16 \\
Clos\_compact                        & 64  & 28 & 4.00 & 8.00  & 16 \\
Fat-Tree                             & 64  & 80 & 4.00 & 8.00  & 32 \\
BCube                                & 64  & 16 & 4.00 & 8.00  & 16 \\
\midrule
QFly ($k_{\mathrm{ring}} = N/m - 1$) & 128 & 32 & 4.00 & 35/35 & 32 \\
QFly ($k_{\mathrm{ring}} = m/2$)     & 128 & 32 & 4.00 & 20/20 & 32 \\
QFly ($k_{\mathrm{ring}} = k - m$)   & 128 & 32 & 4.00 & 8/8   & 32 \\
Clos\_tight                          & 128 & 88 & 2.00 & 16.00 & 64 \\
Clos\_compact                        & 150 & 40 & 6.00 & 10.00 & 25 \\
Fat-Tree                             & 128 & 80 & 4.00 & 8.00  & 32 \\
BCube                                & 128 & 96 & 1.33 & 8.00  & 96 \\
\bottomrule
\end{tabular}}
\caption{Topology summary across architectures and system scales.
\#QPUs denotes the total number of QPUs supported by the instantiated fabric.
\#Sw.\ is the total number of active switches across all layers.
\#ToRs is the number of top-of-rack (edge) switches.
QPUs/Rack reports the maximum number of QPUs attached to any ToR.
Ports/Sw.\ reports the maximum switch degree; for QFly, input/output ports are
reported separately.
BCube does not have a meaningful rack abstraction, and QPUs/Rack reflects the
effective attachment multiplicity.}
\label{table:QPU_scale_switches_updated}
\end{table}

\paragraph{Fixed BSMs per switch.}
Figure~\ref{fig:workload_scalability} shows the distributed-to-monolithic latency
ratio $\rho_{\mathrm{lat}} = T_{\mathrm{dist}}/T_{\mathrm{mono}}$ when the number
of BSMs is fixed per switch. Results are reported \emph{separately for each
workload}, with each bar representing the average over 50 independent runs.
Across all architectures, the latency ratio generally increases with system
size, as a larger fraction of two-qubit gates must be executed remotely. This effect is most pronounced for the long-range workload
(Fig.~\ref{fig:workload_scalability}c), which places the heaviest demand on the
interconnect, while the nearest-neighbor workload
(Fig.~\ref{fig:workload_scalability}a) remains nearly flat since most operations
can be executed locally.
Under this model, architectures with many switches benefit from a larger
aggregate BSM budget: \texttt{clos\_tight} and Fat-Tree achieve the lowest
latency ratios, with \texttt{clos\_compact} slightly worse. QFly remains competitive but does not dominate, as
its small switch footprint limits the total BSM budget. BCube’s behavior is more
parameter-dependent, influenced by the coherence cutoff and by whether QPUs act
as repeaters, which introduce additional memory and swap overhead. We will explore this later in this section. Note that in some cases increasing the number of QPUs also exposes additional
parallelism that can partially offset the higher gate count, leading to small
local decreases in the ratio.

\begin{figure}
    \centering
    \includegraphics[scale=0.5]{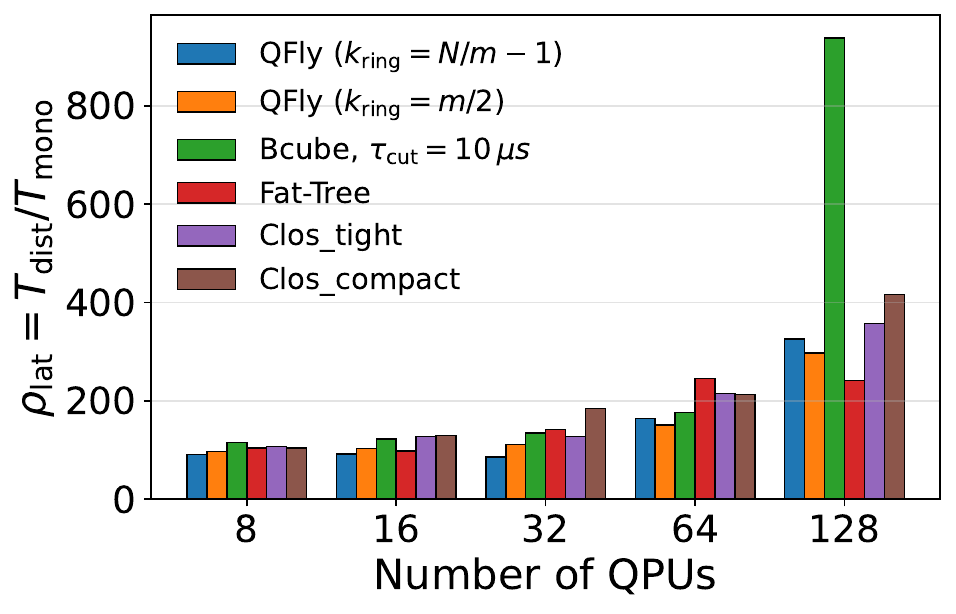}
    \caption{Latency ratio (Distributed / Monolithic) versus number of QPUs under the
\emph{total BSMs budget} resource model for the long-range workload. Bars
show the average ratio across repeated runs for all evaluated architectures.
Lower values indicate better scalability relative to the monolithic baseline.}
    \label{fig:scale_total_BSM}
\end{figure}

\paragraph{Fixed total BSM budget.}
For the fixed-total-BSM model, we focus exclusively on the long-range workload,
as it represents the most communication-intensive setting.
When the global BSM budget is held constant and evenly distributed across
switches, architectures with fewer switches receive a larger per-switch share
of resources (here BSMs), reversing the scaling advantage.
In this regime, QFly performs best overall as shown in figure \ref{fig:scale_total_BSM}, with the fully connected variant
($k_{\mathrm{ring}} = N/m - 1$) achieving the lowest latency ratios across scales.
Even under port-feasible configurations ($k_{\mathrm{ring}} = m/2$), QFly
consistently outperforms Clos and Fat-Tree. In contrast, Clos and Fat-Tree spread
the fixed budget over many more switches, increasing contention; this effect
compounds with their longer multi-stage paths and higher entanglement-generation
latency as shown in Fig. \ref{fig:non_local_per_step}.

\subsubsection{Path Hop-Count Distribution of Non-Local Gates}
\begin{figure}
    \centering
    \includegraphics[scale=0.95]{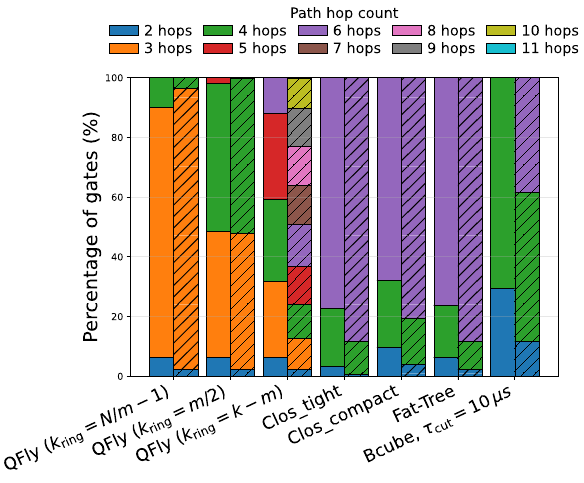}
  \caption{Fraction of non-local gates served over paths of different hop counts
for 32- and 128-QPU configurations across the evaluated architectures. Solid
rectangles correspond to 32 QPUs, while dashed rectangles correspond to 128
QPUs.}

    \label{fig:non_local_per_step}
\end{figure}

We next examine architectural differences by analyzing the hop-count
distribution of paths used to serve non-local gates when the number of BSMs is
fixed per switch. Figure~\ref{fig:non_local_per_step} reports the fraction of
non-local gates as a function of path length for systems with 32 and 128 QPUs
across architectures.

As system size increases from 32 to 128 QPUs, all architectures exhibit a shift
toward longer paths. Notably, in switch-centric architectures such as Fat-Tree
and both Clos variants, a larger fraction of non-local gates are served over
six-hop paths than in BCube at 128 QPUs. Despite this increase in hop count,
Fat-Tree and Clos experience only a modest increase in the distributed-to-
monolithic latency ratio (Fig.~\ref{fig:workload_scalability}c).

In contrast, BCube exhibits a substantially larger latency increase even though
its hop-count distribution is comparable to that
of switch-centric designs. The key reason lies in how hops are realized. In
BCube, paths longer than two hops necessarily traverse intermediate QPUs acting
as repeaters, so each hop incurs quantum-memory loss and requires an additional
Bell-state measurement for entanglement swapping. In our model, this corresponds
to a per-hop memory loss of 3\,dB, compared to 0.3\,dB per switch traversal in
switch-centric architectures. This accumulated memory loss significantly
increases the expected EPR-generation latency and amplifies the performance
penalty at larger scales.

By contrast, switch-centric architectures establish end-to-end entanglement using
a single BSM along the path and avoid intermediate quantum memory, making their
latency substantially less sensitive to increased hop count. While this behavior
is intrinsic to the server-centric design of BCube, it suggests that
BCube-specific scheduling and swap-coordination algorithms could mitigate part of
this overhead.

\subsection{Sensitivity of BCube to communication qubits coherence time}

Next, we study the impact of the cutoff time parameter $\tau_{\mathrm{cut}}$ on
distributed circuit execution latency by comparing different quantum
data-center architectures relative to the \emph{BCube} architecture. As discussed
earlier, entanglement-swap
scheduling protocol and the duration for which partially generated
entanglement can be retained in memory before being discarded can affect the performance of server-centric architectures. Throughout this
section, we use the terms \emph{coherence time} and \emph{cutoff time}
interchangeably, as both refer to this memory-lifetime constraint. 

In this experiment, we fix the circuit and network configuration (long-range
workload, fixed number of QPUs, communication qubits, and BSM allocation), set
the cutoff time $\tau_{\mathrm{cut}}$ in the BCube network model, simulate
distributed circuit execution, and record the resulting execution latency.
For BCube, we use the parallel swap scheduling protocol proposed in \cite{pouryousef2024minimal}.

\begin{figure}
    \centering
    \includegraphics[scale=0.46]{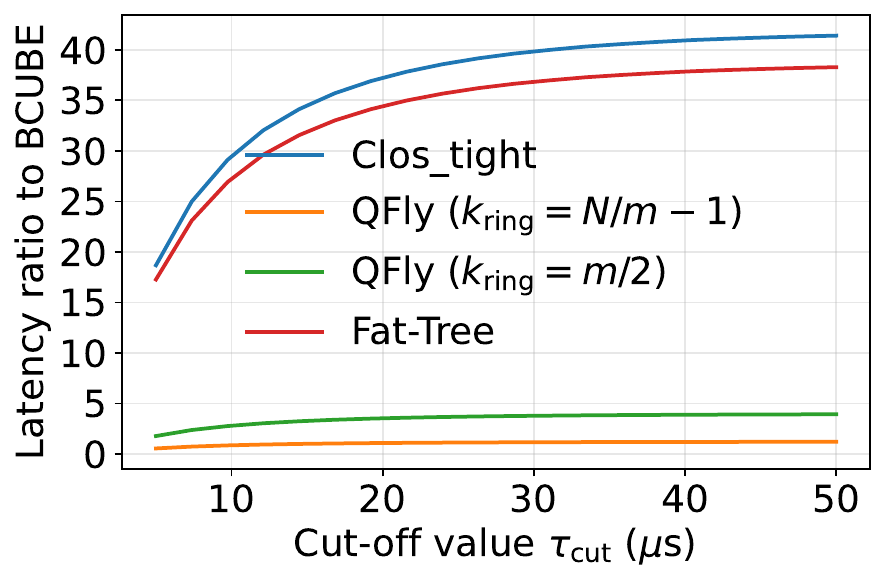}
   \caption{Latency ratio of switch-centric architectures to BCube as a function of
the cut-off value $\tau_{\mathrm{cut}}$ under the parallel swap scheduling
protocol for 128 QPUs, 5 communication qubits per QPU and total BSM budget of 100 evenly distributed among switches.}
    \label{fig:cutoff_time_BCUBE_ratio}
\end{figure}

Figure~\ref{fig:cutoff_time_BCUBE_ratio} shows the latency ratio of each
architecture relative to BCube as a function of the communication-qubit memory
cutoff value $\tau_{\mathrm{cut}}$.
The cutoff value controls how long an entangled state generated on any
\emph{entanglement segment}—defined here as a single QPU--switch--QPU link—can be
stored in quantum memory before being discarded if entanglement on the remaining
segments of the path has not yet been established.
In BCube, non-local operations often require multi-hop paths composed of multiple
such segments (e.g., a four-hop path consisting of two QPU--switch--QPU
segments), where entanglement generation on different segments proceeds in
parallel and may complete at different times.
A larger $\tau_{\mathrm{cut}}$ allows early-completing segments to wait longer
for the remaining segments to succeed, increasing the probability that all
segments are simultaneously available and reducing the expected
entanglement-generation latency.
As a result, BCube performance improves steadily as $\tau_{\mathrm{cut}}$
increases, leading to an increasing latency ratio of switch-centric
architectures relative to BCube.

\begin{figure}
    \centering
    \includegraphics[scale=0.46]{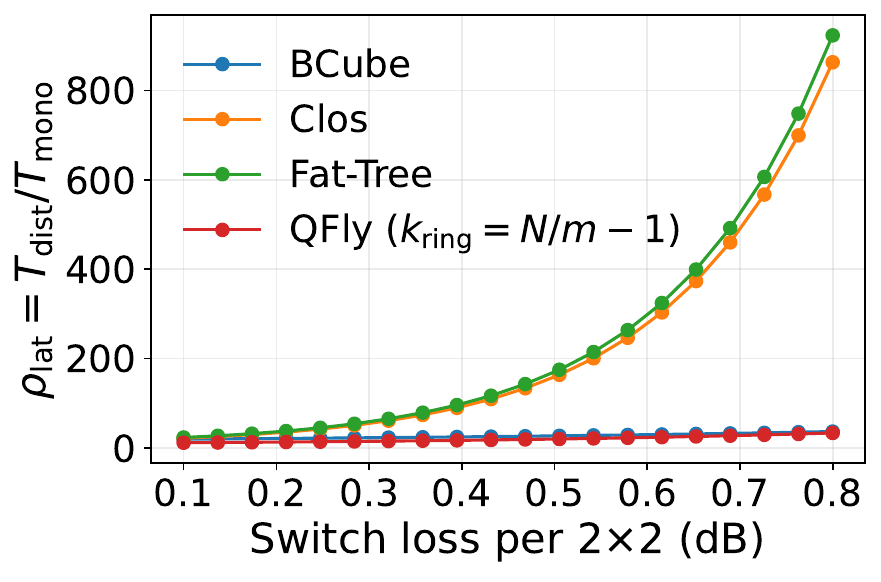}
    \caption{Latency ratio versus per-switch insertion loss, showing higher sensitivity for deeper switch-centric architectures (Clos and Fat-Tree) compared to QFly and BCube.}
\label{fig:ratio_to_swtich_loss}
\end{figure}

\subsection{Switch Improvements vs. Circuit Execution Latency}
\label{sec:loss-delay-affect}

We now report the results of the switch-loss sensitivity experiment. For each
architecture, we fix the workload (\emph{long\_range}) and all non-loss
parameters (including a fixed switch reconfiguration duration), and directly
sweep the switch insertion loss per $2\times2$ element, $\ell_{2\times2}$, over a
practical range. For each loss configuration, we run the distributed circuit
execution, evaluate the resulting end-to-end execution latency using our
physical-layer entanglement timing model, and report the latency ratio
(distributed / monolithic) averaged over repeated runs.

\paragraph{Observations.}
Figure~\ref{fig:ratio_to_swtich_loss} shows that increasing switch insertion loss
monotonically worsens the latency ratio for all architectures. However, the
degradation is substantially steeper for Clos and Fat-Tree than for QFly and
BCube. The underlying reason is architectural: deeper switch-centric designs
(Clos and Fat-Tree) realize remote entanglement over paths that traverse a larger
number of optical switches, so increasing $\ell_{2\times2}$ accumulates loss more
aggressively along the path, reducing the success probability of each attempt
and inflating the expected EPR-generation latency.

In contrast, QFly typically achieves remote connectivity with fewer switch
traversals, making it less sensitive to the same per-switch loss increase. BCube
is also comparatively less sensitive in this sweep because some intermediate
nodes on its entanglement paths are \emph{QPUs acting as repeaters}, which we do
not model as contributing switch insertion loss; thus, raising $\ell_{2\times2}$
does not penalize those hops. (We keep the repeater memory-loss parameter fixed
throughout this experiment, so the observed trends isolate the impact of switch
insertion loss.)

\section{Conclusion and Discussion}
\label{sec:conclusion}
We presented a systematic benchmarking study of QDC
architectures for DQC, quantifying how topology,
scheduling, and physical-layer constraints jointly determine end-to-end circuit
execution latency. Across multiple QFly variants, BCube, Clos, and Fat-Tree, and
under realistic entanglement-generation models and diverse workloads, our
results show that distributed quantum performance cannot be inferred from
topology alone: it emerges from the interaction of multiple factors including path structure, switching
resources, and coherence-limited entanglement dynamics.

Our evaluation exposes clear architectural trade-offs. Switch-centric fabrics
(QFly/Clos/Fat-Tree) avoid using QPUs as repeaters, reducing sensitivity to
communication-qubit coherence cutoffs, but they incur switch loss on every hop
and can become bottlenecked by contention for shared BSM resources—especially
in deeper hierarchies with longer end-to-end paths. In contrast, the
server-centric BCube design can exploit shorter effective paths and QPU-assisted
swapping, but its latency grows with scale when repeater chains become longer
and performance becomes sensitive to finite retry windows and memory-induced
loss on intermediate QPUs. Across all architectures, hop count is a primary
driver of both optical-loss accumulation and expected EPR generation time, and
thus a key determinant of scalability on communication-heavy workloads.

Our sensitivity experiments further show that hardware improvements yield
architecture-dependent benefits. Increasing switch insertion loss degrades
latency ratios for all designs, but the penalty is substantially steeper for
switch-rich, multi-stage fabrics (Clos/Fat-Tree) because remote entanglement
paths traverse more switches; QFly, with fewer switch traversals, is less
sensitive, and BCube is comparatively less affected because intermediate QPU
repeaters are not modeled as incurring switch insertion loss in those hops (with
memory-loss held fixed in that sweep).

These results suggest several directions for co-design. First, topology should
be chosen and parameterized jointly with BSM placement and scheduling: path
diversity only translates into latency gains when the scheduling policy can
exploit it under realistic BSM contention. Second, physical-layer control knobs
(coherence windows, retry policies, and switch configuration strategies) should
be exposed to the scheduler and tuned to workload structure, since long-range
workloads amplify sensitivity to hop count, loss, and shared resources. Finally,
future QDC platforms will likely require cross-layer optimization that couples
architectural design (diameter and switch footprint), orchestration (routing and
contention-aware scheduling), and photonic hardware choices (loss, switching
latency, and detector/BSM provisioning) to achieve robust scaling for DQC.

Overall, our study provides quantitative guidance for designing scalable quantum
data centers and establishes a benchmarking foundation for evaluating emerging
QDC architectures and future fault-tolerant distributed execution stacks.

\section{Acknowledgements}
The authors acknowledge insightful discussions with Amin Taherkhani, Jiapeng Zhao, Narges Alavisamani, and Ramin Ayanzadeh.


\bibliographystyle{ACM-Reference-Format}
\bibliography{refs}

@inproceedings{baker2020time,
  title={Time-sliced quantum circuit partitioning for modular architectures},
  author={Baker, Jonathan M and Duckering, Casey and Hoover, Alexander and Chong, Frederic T},
  booktitle={Proceedings of the 17th ACM International Conference on Computing Frontiers},
  pages={98--107},
  year={2020}
}

@inproceedings{burt2024generalised,
  title={Generalised circuit partitioning for distributed quantum computing},
  author={Burt, Felix and Chen, Kuan-Cheng and Leung, Kin K},
  booktitle={2024 IEEE International Conference on Quantum Computing and Engineering (QCE)},
  volume={2},
  pages={173--178},
  year={2024},
  organization={IEEE}
}

@article{beukers2024remote,
  title={Remote-entanglement protocols for stationary qubits with photonic interfaces},
  author={Beukers, Hans KC and Pasini, Matteo and Choi, Hyeongrak and Englund, Dirk and Hanson, Ronald and Borregaard, Johannes},
  journal={PRX Quantum},
  volume={5},
  number={1},
  pages={010202},
  year={2024},
  publisher={APS}
}

@article{spanke1987n,
  title={N-stage planar optical permutation network},
  author={Spanke, Ron A and Benes, VE},
  journal={Applied Optics},
  volume={26},
  number={7},
  pages={1226--1229},
  year={1987},
  publisher={Optical Society of America}
}

@article{main2025distributed,
  title={Distributed quantum computing across an optical network link},
  author={Main, D and Drmota, P and Nadlinger, DP and Ainley, EM and Agrawal, A and Nichol, BC and Srinivas, R and Araneda, G and Lucas, DM},
  journal={Nature},
  pages={1--6},
  year={2025},
  publisher={Nature Publishing Group UK London}
}

@article{sakuma2024optical,
  title={An optical interconnect for modular quantum computers},
  author={Sakuma, Daisuke and Taherkhani, Amin and Tsuno, Tomoki and Sasaki, Toshihiko and Shimizu, Hikaru and Teramoto, Kentaro and Todd, Andrew and Ueno, Yosuke and Hajdu{\v{s}}ek, Michal and Ikuta, Rikizo and others},
  journal={arXiv preprint arXiv:2412.09299},
  year={2024}
}

@article{sinclair2025fault,
  title={Fault-tolerant optical interconnects for neutral-atom arrays},
  author={Sinclair, Josiah and Ramette, Joshua and Grinkemeyer, Brandon and Bluvstein, Dolev and Lukin, Mikhail D and Vuleti{\'c}, Vladan},
  journal={Physical Review Research},
  volume={7},
  number={1},
  pages={013313},
  year={2025},
  publisher={APS}
}

@article{cirac1999distributed,
  title={Distributed quantum computation over noisy channels},
  author={Cirac, J Ignacio and Ekert, AK and Huelga, Susana F and Macchiavello, Chiara},
  journal={Physical Review A},
  volume={59},
  number={6},
  pages={4249},
  year={1999},
  publisher={APS}
}

@inproceedings{zhang2025switchqnet,
  title={SwitchQNet: Optimizing Distributed Quantum Computing for Quantum Data Centers with Switch Networks},
  author={Zhang, Hezi and Xu, Yiran and Hu, Haotian and Yin, Keyi and Shapourian, Hassan and Zhao, Jiapeng and Kompella, Ramana Rao and Nejabati, Reza and Ding, Yufei},
  booktitle={Proceedings of the 52nd Annual International Symposium on Computer Architecture},
  pages={1449--1463},
  year={2025}
}

@article{shapourian2025quantum,
  title={Quantum data center infrastructures: A scalable architectural design perspective},
  author={Shapourian, Hassan and Kaur, Eneet and Sewell, Troy and Zhao, Jiapeng and Kilzer, Michael and Kompella, Ramana and Nejabati, Reza},
  journal={arXiv preprint arXiv:2501.05598},
  year={2025}
}

@inproceedings{popa2010cost,
  title={A cost comparison of datacenter network architectures},
  author={Popa, Lucian and Ratnasamy, Sylvia and Iannaccone, Gianluca and Krishnamurthy, Arvind and Stoica, Ion},
  booktitle={Proceedings of the 6th International COnference},
  pages={1--12},
  year={2010}
}

@book{liu2013data,
  title={Data center networks: Topologies, architectures and fault-tolerance characteristics},
  author={Liu, Yang and Muppala, Jogesh K and Veeraraghavan, Malathi and Lin, Dong and Hamdi, Mounir},
  year={2013},
  publisher={Springer Science \& Business Media}
}

@article{xia2016survey,
  title={A survey on data center networking (DCN): Infrastructure and operations},
  author={Xia, Wenfeng and Zhao, Peng and Wen, Yonggang and Xie, Haiyong},
  journal={IEEE communications surveys \& tutorials},
  volume={19},
  number={1},
  pages={640--656},
  year={2016},
  publisher={IEEE}
}

@article{clos1953study,
  title   = {A Study of Non-Blocking Switching Networks},
  author  = {Clos, Charles},
  journal = {Bell System Technical Journal},
  volume  = {32},
  number  = {2},
  pages   = {406--424},
  year    = {1953}
}

@inproceedings{greenberg2009vl2,
  title     = {{VL2}: A Scalable and Flexible Data Center Network},
  author    = {Greenberg, Albert and Hamilton, James and Jain, Navendu and Kandula, Srikanth and Kim, Changhoon and Lahiri, Parantapa and Maltz, David A. and Patel, Parveen and Sengupta, Sudipta},
  booktitle = {Proceedings of the ACM SIGCOMM Conference},
  year      = {2009},
  pages     = {51--62}
}

@inproceedings{singh2015jupiter,
  title     = {Jupiter Rising: A Decade of Clos Topologies and Centralized Control in Google’s Datacenter Network},
  author    = {Singh, Arjun and Ong, Joon and Agarwal, Amit and Anderson, Glen and Armistead, Ashby and Bannon, Roy and Boving, Seb and Desai, Gaurav and Felderman, Bob and Germano, Paul and others},
  booktitle = {Proceedings of the ACM SIGCOMM Conference},
  year      = {2015},
  pages     = {183--197}
}

@article{al2008scalable,
  title={A scalable, commodity data center network architecture},
  author={Al-Fares, Mohammad and Loukissas, Alexander and Vahdat, Amin},
  journal={ACM SIGCOMM computer communication review},
  volume={38},
  number={4},
  pages={63--74},
  year={2008},
  publisher={ACM New York, NY, USA}
}

@article{kaur2025optimized,
  title={Optimized quantum circuit partitioning across multiple quantum processors},
  author={Kaur, Eneet and Pouryousef, Shahrooz and Shapourian, Hassan and Zhao, Jiapeng and Kilzer, Michael and Kompella, Ramana and Nejabati, Reza},
  journal={IEEE Transactions on Quantum Engineering},
  year={2025},
  publisher={IEEE}
}

@article{kim2008technology,
  title={Technology-driven, highly-scalable dragonfly topology},
  author={Kim, John and Dally, Wiliam J and Scott, Steve and Abts, Dennis},
  journal={ACM SIGARCH Computer Architecture News},
  volume={36},
  number={3},
  pages={77--88},
  year={2008},
  publisher={ACM New York, NY, USA}
}

@inproceedings{guo2009bcube,
  title={BCube: a high performance, server-centric network architecture for modular data centers},
  author={Guo, Chuanxiong and Lu, Guohan and Li, Dan and Wu, Haitao and Zhang, Xuan and Shi, Yunfeng and Tian, Chen and Zhang, Yongguang and Lu, Songwu},
  booktitle={Proceedings of the ACM SIGCOMM 2009 conference on Data communication},
  pages={63--74},
  year={2009}
}

@article{watkins2024high,
  title={A high performance compiler for very large scale surface code computations},
  author={Watkins, George and Nguyen, Hoang Minh and Watkins, Keelan and Pearce, Steven and Lau, Hoi-Kwan and Paler, Alexandru},
  journal={Quantum},
  volume={8},
  pages={1354},
  year={2024},
  publisher={Verein zur F{\"o}rderung des Open Access Publizierens in den Quantenwissenschaften}
}

@inproceedings{guo2008dcell,
  title={Dcell: a scalable and fault-tolerant network structure for data centers},
  author={Guo, Chuanxiong and Wu, Haitao and Tan, Kun and Shi, Lei and Zhang, Yongguang and Lu, Songwu},
  booktitle={Proceedings of the ACM SIGCOMM 2008 conference on Data communication},
  pages={75--86},
  year={2008}
}

@article{leiserson1985fattree,
  title   = {Fat-Trees: Universal Networks for Hardware-Efficient Supercomputing},
  author  = {Leiserson, Charles E.},
  journal = {IEEE Transactions on Computers},
  volume  = {C-34},
  number  = {10},
  pages   = {892--901},
  year    = {1985}
}

@inproceedings{alfares2008fattree,
  title     = {A Scalable, Commodity Data Center Network Architecture},
  author    = {Al-Fares, Mohammad and Loukissas, Alexander and Vahdat, Amin},
  booktitle = {Proceedings of the ACM SIGCOMM Conference},
  year      = {2008},
  pages     = {63--74}
}

@article{horsman2012surface,
  title={Surface code quantum computing by lattice surgery},
  author={Horsman, Dominic and Fowler, Austin G and Devitt, Simon and Van Meter, Rodney},
  journal={New Journal of Physics},
  volume={14},
  number={12},
  pages={123011},
  year={2012},
  publisher={IOP Publishing}
}

@article{cuomo2023optimized,
  title={Optimized compiler for distributed quantum computing},
  author={Cuomo, Daniele and Caleffi, Marcello and Krsulich, Kevin and Tramonto, Filippo and Agliardi, Gabriele and Prati, Enrico and Cacciapuoti, Angela Sara},
  journal={ACM Transactions on Quantum Computing},
  volume={4},
  number={2},
  pages={1--29},
  year={2023},
  publisher={ACM New York, NY}
}

@inproceedings{pouryousef2024minimal,
  title={Minimal Protocols for Entanglement Distribution with Finite Memory Coherence Time},
  author={Pouryousef, Shahrooz and Shapourian, Hassan and Towsley, Don},
  booktitle={2024 International Conference on Quantum Communications, Networking, and Computing (QCNC)},
  pages={150--159},
  year={2024},
  organization={IEEE}
}

@article{li2025optimising,
  title={Optimising entanglement distribution policies under classical communication constraints assisted by reinforcement learning},
  author={Li, Jan and Coopmans, Tim and Emonts, Patrick and Goodenough, Kenneth and Tura, Jordi and van Nieuwenburg, Evert},
  journal={Machine Learning: Science and Technology},
  volume={6},
  number={3},
  pages={035024},
  year={2025},
  publisher={IOP Publishing}
}

@article{monroe2014large,
  title={Large-scale modular quantum-computer architecture with atomic memory and photonic interconnects},
  author={Monroe, Christopher and Raussendorf, Robert and Ruthven, Alex and Brown, Kenneth R and Maunz, Peter and Duan, L-M and Kim, Jungsang},
  journal={Physical Review A},
  volume={89},
  number={2},
  pages={022317},
  year={2014},
  publisher={APS}
}

@article{jiang2007distributed,
  title={Distributed quantum computation based on small quantum registers},
  author={Jiang, Liang and Taylor, Jacob M and S{\o}rensen, Anders S and Lukin, Mikhail D},
  journal={Physical Review A—Atomic, Molecular, and Optical Physics},
  volume={76},
  number={6},
  pages={062323},
  year={2007},
  publisher={APS}
}

@article{keskin2025lattice,
  title={Lattice Surgery Aware Resource Analysis for the Mapping and Scheduling of Quantum Circuits for Scalable Modular Architectures},
  author={Keskin, Batuhan and Afradi, Cameron and Lovis, Sylvain and Palesi, Maurizio and Escofet, Pau and Almudever, Carmen G and Charbon, Edoardo},
  journal={arXiv preprint arXiv:2511.21885},
  year={2025}
}

@ARTICLE{KL_algorithm,
  author={Kernighan, B. W. and Lin, S.},
  journal={The Bell System Technical Journal}, 
  title={An efficient heuristic procedure for partitioning graphs}, 
  year={1970},
  volume={49},
  number={2},
  pages={291-307},
  doi={10.1002/j.1538-7305.1970.tb01770.x}}

\end{document}